\documentclass[12pt]{article}
\usepackage{graphicx}
\usepackage{epsfig}
\usepackage{amsbsy}

\begin{document}
\baselineskip 10mm

\centerline{\large \bf Spontaneous Regeneration}
\centerline{\large \bf of an Atomically Sharp 
Graphene/Graphane Interface}
\centerline{\large \bf under Thermal Disordering}

\vskip 6mm

\centerline{L. A. Openov$^{*}$ and A. I. Podlivaev}

\vskip 4mm

\centerline{\it Moscow Engineering Physics Institute (State
University), 115409 Moscow, Russia}

\vskip 2mm

$^{*}$ E-mail: LAOpenov@mephi.ru

\vskip 8mm

\centerline{\bf ABSTRACT}

The smearing of the graphene/graphane interface due to the thermally activated migration of hydrogen
atoms is studied1 by the molecular dynamics method. Contrary to expectations, it is found that the
fast spontaneous regeneration of this interface occurs even at a sufficiently high temperature
$T\approx 1500$ K. As a result, the average width of the disordered region does not exceed the length
of a C-C bond, i.e., the interface remains almost atomically sharp. The cause of this effect appears
to be the specific shape of the potential relief of the system, namely, the significant difference
between the heights of the energy barriers for the direct and inverse migrations of hydrogen atoms.
A simple model that makes it possible to obtain the temperature dependence of the equilibrium
distribution function of typical atomic configurations, to estimate the typical time of establishing
the equilibrium state, and thereby to quantitatively describe the results of the computer experiment
is presented.

\vskip 5mm


\newpage

The existence of graphane, i.e., a graphene monolayer completely saturated by hydrogen from both
sides, was recently predicted in theoretical work [1]. The experimental synthesis of graphane [2]
put forth the problem of its possible applications. In contrast to graphene, graphane is an insulator
and can be used in nanoelectronics in combination with graphene [3]. For example, various
nanoelectronic devices can be manufactured by the selective sorption of hydrogen on graphene or
graphene nanoribbons. At first glance, it seems that the operation temperatures of such devices
should be very low, because the thermally activated migration of hydrogen atoms through
graphene/graphane interfaces gives rise to the fast smearing of these interfaces and an uncontrolled
change in the electrophysical characteristics of a device. Below, we show that this is not necessary:
the computer experiment on the numerical simulation of the dynamics of the graphene/graphane system
demonstrates that the interface resists thermal disordering and remains atomically sharp even at
high temperatures. This effect is not only of fundamental interest, but also important for
applications.

Our first aim was to determine the temperature dependence of the smearing rate of the
graphene/graphane interface. The initial graphene sample was simulated by a 88-atom fragment of the
hexagonal carbon monolayer with edges passivated by hydrogen in order to saturate the dangling bonds
of  $sp$-hybridized carbon atoms to weaken the effects of finite sizes (the number of passivating
hydrogen atoms is 26). One half of this sample was transformed to graphane by alternating the
bonding of one hydrogen atom to each of the 44 carbon atoms from both sides of the plane of the
initial monolayer (i.e., the orientation of each hydrogen atom is determined by a graphene
sublattice (of two equivalent sublattices) to which the nearest carbon atom belongs. As a result,
we obtain the C$_{88}$H$_{70}$ cluster presented in Fig. 1 (configuration {}``A'' in Fig. 2).

To simulate the thermally activated migration of hydrogen through the graphene/graphane interface,
we used the molecular dynamics method [4-6]. At the initial time, random velocities and
displacements were assigned to all of the atoms so that the momentum and angular momentum of the
cluster were zero. Then, the forces acting on the atoms were calculated. The classical Newtonian
equations of motion were numerically integrated using the velocity Verlet method with a time
step of $t_{0}=2.72\times 10^{-16}$ s. The total energy of the system (the sum of the potential and
kinetic energies) in the simulation process is conserved, which corresponds to a microcanonical
ensemble (the system is thermally isolated from the environment) [4-6]. In this case, the
{}``dynamic'' temperature $T$ is a measure of the energy of the relative motion of atoms and is
calculated from the formula [5]
$\langle E_{\textrm{kin}} \rangle=\frac{1}{2}k_{B}T(3n-6)$,
where $\langle E_{\textrm{kin}} \rangle$ is the time-average kinetic energy of the system, $k_B$ is
the Boltzmann constant, and $n$ is the number of atoms in the system ($n$ = 158 in our case).
To calculate the interatomic interaction forces, a nonorthogonal tight binding model [8] modified
as compared to [7] was used. This model is a reasonable compromise between stricter ab initio
methods and too simplified classical potentials of the interatomic interaction. It reasonably
describes both small carbon (e.g., fullerenes [8]) and hydrocarbon (e.g., cubane C$_8$H$_8$ [9, 10])
clusters and macroscopic systems [8] and requires much less computer resources than ab initio
methods; for this reason, it allows to study the evolution of the system of $\sim 100$ atoms for
a time of $\sim 1$ ns sufficient for the collection of necessary statistics.

At $T=$ 2000 - 2500 K in time $< 10$ ps, the interface is completely smeared due to the migration of
a large number of hydrogen atoms from graphane to graphene and/or their desorption. However, as the
temperature is decreased to $T=$ 1500 - 1800 K, we observed the following picture. The migration of
one hydrogen atom by the distance of a C-C bond (configuration {}``B'' in Fig. 2 and similar
configurations formed after the migration of other boundary hydrogen atoms by the distance of a C-C
bond) is usually followed by a fast (in a time of $\sim 1$ ps) hop of this atom to the initial
position (i.e., the return to configuration {}``A'' occurs), whereas its repeated migration to the
graphene region (configuration {}``C'' in Fig. 2 and similar configurations) occurs very rarely.
After the first elementary migration act, the liberated site is sometimes occupied by a hydrogen
atom from the other sublattice (configuration {}``D'' in Fig. 2 and similar configurations), which
soon (in a time of $\sim 1$ ps) returns to its position, i.e., configuration {}``B'' is recovered;
then, it is transformed to configuration {}``A''. Sometimes, we observed the regeneration of the
graphene/graphane interface even after a much more complex sequence of hops of several hydrogen
atoms belonging to different sublattices. At $T=$ 1700 - 1800 K, 10-15 complete recoveries of the
disordered interface occurred during the simulation time ($\sim 0.1$ ns); after that, either the
desorption of one hydrogen atom or molecule occurred or the width of the disordered region reached
a length of several C-C bonds; i.e., the smearing of the interface became irreversible. A further
decrease in the temperature results in a strong increase in the typical onset time of the
disordering of the interface; as a result, the mean time interval between two successive recoveries
and, therefore, the time required for the irreversible smearing of this interface, increased.

To determine the cause of the thermal stability of the graphene/graphane interface, we
examine the form of the hypersurface of the potential energy of the system,
$E_{\textrm{pot}}$, as a function of the coordinates of the constituent atoms and obtain
the heights of the energy barriers separating atomic configurations {}``A'', {}``B'',
{}``C'', and {}``D'' shown in Fig. 2 (the calculation method was presented in more detail
in [4, 5, 11, 12]). Figure 3 shows the profile of $E_{\textrm{pot}}$  along the reaction
coordinate passing through configurations {}``A'', {}``B'', and {}``D''. It is seen that
for the {}``A''$\leftrightarrow${}``B'' and {}``B''$\leftrightarrow${}``D'' transitions,
the heights $U_{\textrm{AB}}$ and $U_{\textrm{BD}}$ of the barriers preventing
disordering are larger than the heights $U_{\textrm{BA}}$ and $U_{\textrm{DB}}$,
respectively, of the barriers preventing the return of the system to the initial state
after the migration of one and two hydrogen atoms. For the transitions between
configurations {}``B'' and {}``C'', the barrier for the transition to configuration
{}``C'' that is farther from the initial configuration is also higher than the barrier for
the inverse transition; i.e., $U_{\textrm{BC}} > U_{\textrm{CB}}$. The calculated
heights of the barriers are $U_{\textrm{AB}}=0.96$ eV, $U_{\textrm{BA}}=0.35$ eV,
$U_{\textrm{BD}}=0.50$ eV, $U_{\textrm{DB}}=0.39$ eV, $U_{\textrm{BC}}=0.81$ eV,
$U_{\textrm{CB}}=0.62$ eV. According to the Arrhenius formula
\begin{equation}
P_{ij}(T)=A_{ij}\times\exp\left(-\frac{U_{ij}}{k_{B}T}\right)~,
\label{1}
\end{equation}
where $P_{ij}$ is the probability of the $i\rightarrow j$ transition per unit time,
$U_{ij}$ is the height of the barrier between two atomic configurations $i$ and $j$,
and $A_{ij}$ is the frequency factor with the dimension s$^{-1}$; under the conditions
$U_{ij} > U_{ji}$ and $k_B T << U_{ji}$, the system is more often in configuration
$i$ than in configuration $j$ (if one of the frequency factors is not much smaller or
much larger than the other).

Since $U_{\textrm{BC}} > U_{\textrm{BD}} > U_{\textrm{BA}}$, it is clear why at
$k_B T << U_{\textrm{BA}}$, first, the system almost always returns to the initial
state after the migration of one hydrogen atom; second, the subsequent migration of the
hydrogen atom in the other sublattice occurs much more rarely; and, third, the
displacement of an atom from the interface by the distance of two C-C bonds has a low
probability. The same relations exist between the heights of the barriers separating the
configurations formed after the migration of the other four boundary hydrogen atoms of
our model system (see Fig. 2) and the subsequent migration of the corresponding atoms of
the other sublattice. Thus, the physical cause of the thermal stability of the
graphene/graphane interface is the specific shape of the potential relief of the system,
namely, the significant difference between the heights of the energy barriers for direct
and inverse hops of the hydrogen atoms, see Figs. 2 and 3. It is worth noting that the
potential relief for the simultaneous migration of several hydrogen atoms is much more
complex and, at sufficiently strong disordering, the barrier for the inverse migration
of a certain atom is sometimes higher than the barrier overcome by this atom in the
path to a given configuration. In particular, this concerns migration along the
interface. As a result, the system can be for a long time in a state strongly different
from the initial state. However, the migration of one atom by the distance of two or
more C-C bonds from the interface and the simultaneous migration of several atoms occur
more rarely with a decreasing temperature. In our computer experiment for $T=1500$ K,
such configurations do not appear in a time of about 1 ns, which corresponds to
3 x 10$^6$ steps of molecular dynamics. With a further decrease in temperature, an
exponential increase is expected in the time interval in which only minimally disordered
configurations {}``B'' and {}``D'' are formed (and rapidly {}``healed'').

To estimate the thermally equilibrium distribution function $f_i$ of configurations
{}``A'', {}``B'', and {}``D'' most often observed in the simulation, we used the
chemical kinetic equations based on the following model. Let us consider the statistical
ensemble of a large number of graphene/graphane systems (C$_{88}$H$_{70}$ clusters in
our case). At the initial time, all of them are in configuration {}``A''. The
probabilities $f_{\textrm{A}}(t)$, $f_{\textrm{B}}(t)$, and $f_{\textrm{D}}(t)$ of
finding the system in configurations {}``A'', {}``B'', and {}``D'', respectively,
at time $t$ satisfy the system of differential equations
\begin{eqnarray}
&&\frac{df_{\textrm{A}}(t)}{dt}=
P_{\textrm{BA}}f_{\textrm{B}}(t)-P_{\textrm{AB}}f_{\textrm{A}}(t)~,\nonumber \\
&&\frac{df_{\textrm{B}}(t)}{dt}=
P_{\textrm{AB}}f_{\textrm{A}}(t)-P_{\textrm{BA}}f_{\textrm{B}}(t)+
2P_{\textrm{DB}}f_{\textrm{D}}(t)-2P_{\textrm{BD}}f_{\textrm{B}}(t)~,\nonumber \\
&&2\frac{df_{\textrm{D}}(t)}{dt}=
2P_{\textrm{BD}}f_{\textrm{B}}(t)-2P_{\textrm{DB}}f_{\textrm{D}}(t)
\label{2}
\end{eqnarray}
with the initial condition
\begin{equation}
f_{\textrm{A}}(0)=1,~f_{\textrm{B}}(0)=0,~f_{\textrm{D}}(0)=0,
\label{3}
\end{equation}
Here, the conditional transition probabilities $P_{ij}$ are given by Eq. (1) and we take
into account only the {}``A'' $\leftrightarrow$ {}``B'' and
{}``B'' $\leftrightarrow$ {}``D'' transitions (i.e., neglect the transitions to more
disordered configurations) and the existence of two equivalent {}``D'' configurations,
see Fig. 2. Note that $f_{\textrm{A}}(t)+f_{\textrm{B}}(t)+2f_{\textrm{D}}(t)=1$ at any
time according to Eqs. (2) and (3). In the steady (thermodynamically equilibrium) state,
$df_i(t)/dt=0$ for all of the configurations $i=$ A, B, and D; for this reason, we obtain
the system of linear homogeneous equations for equilibrium (at $t\rightarrow\infty$)
values $f_i$ from Eqs. (2) and (3); the solution of this system has the form
\begin{eqnarray}
&&f_{\textrm{A}}=1-\frac{P_{\textrm{AB}}(P_{\textrm{DB}}+2P_{\textrm{BD}})}
{P_{\textrm{AB}}(P_{\textrm{DB}}+2P_{\textrm{BD}})+P_{\textrm{BA}}P_{\textrm{DB}}}~,\nonumber \\
&&f_{\textrm{B}}=\frac{P_{\textrm{AB}}P_{\textrm{DB}}}
{P_{\textrm{AB}}(P_{\textrm{DB}}+2P_{\textrm{BD}})+P_{\textrm{BA}}P_{\textrm{DB}}}~,\nonumber \\
&&f_{\textrm{D}}=\frac{P_{\textrm{AB}}P_{\textrm{BD}}}
{P_{\textrm{AB}}(P_{\textrm{DB}}+2P_{\textrm{BD}})+P_{\textrm{BA}}P_{\textrm{BD}}}~.
\label{4}
\end{eqnarray}
Since the height $U_{\textrm{AB}}$ is much larger than the heights of all other barriers,
according to Eqs. (1) and (4), $f_{\textrm{A}}$ at $T < 2000$ K is close to unity, while
$f_{\textrm{B}} << 1$ è $f_{\textrm{D}} << 1$.

To determine $f_{\textrm{A}}$, $f_{\textrm{B}}$, and $f_{\textrm{D}}$, it is necessary to
know all of the frequency factors $A_{ij}$ in Eqs. (4). Their calculation is a difficult
problem. Since numerous (about 100) transitions between configurations {}``A'' and {}``B''
were observed in the simulation of the dynamics of the disordering of the
graphene/graphane interface, we collected a lot of statistics and directly determined the
frequency factors of these transitions from the straight line approximation of the
calculated dependences of the transition times
$\tau_{\textrm{AB}}=P_{\textrm{AB}}^{-1}$ and
$\tau_{\textrm{BA}}=P_{\textrm{BA}}^{-1}$ on the inverse temperature using Eq. (1). We
obtain $A_{\textrm{AB}}\approx 2\times 10^{14}$ s$^{-1}$ and
$A_{\textrm{BA}}\approx 5\times 10^{13}$ s$^{-1}$ (the $U_{\textrm{AB}}$ and
$U_{\textrm{BA}}$ values also determined in these calculations coincide within the
statistical errors with the values presented above, which were obtained by a
fundamentally different method). Estimating the frequency factors
$A_{\textrm{BD}}$ and $A_{\textrm{DB}}$ as $10^{14}$ s$^{-1}$, we calculated the
distribution function over the configurations for several temperatures for which the
simulation was performed. For example, for $T=1800$ K, we obtained
$f_{\textrm{A}}=0.866$, $f_{\textrm{B}}=0.068$, and $f_{\textrm{D}}=0.033$. These values
are in good agreement (taking into account rough approximations) with the values
$f_{\textrm{A}}=0.814$, $f_{\textrm{B}}=0.119$, and $f_{\textrm{D}}=0.032$, which were
determined as the relative time intervals during which the system was in the
corresponding states in the simulation of its time evolution (the total number of
molecular dynamics steps was more than 200 000; the relative weight of other, more
disordered configurations was 0.003). This agreement between the statistical and dynamic
data is a consequence of ergodicity (the mean value over the ensemble of the systems is
equal to the average value over the trajectory of one system). The typical time $\tau$
of establishing the equilibrium state can be determined exactly from Eqs. (2) and (3),
but the analytical solution is lengthy and we present only the estimate of
$\tau^{-1} \sim max\{P_{\textrm{AB}},~P_{\textrm{BA}},~P_{\textrm{BD}},~P_{\textrm{DB}}\}$.
In view of the relation
$P_{\textrm{BA}},~P_{\textrm{DB}} >> P_{\textrm{AB}},~P_{\textrm{BD}}$
at a fixed temperature, this time is primarily determined by the
{}``B''$\rightarrow$"A" and {}``D''$\rightarrow$"B" transition rates. For example, for
room temperature, we obtain $\tau\sim 10$ ns.

To conclude, we emphasize that the frequency factor $A_{\textrm{AB}}$ was numerically
determined for a small model system with the length of the graphene/graphane interface
$L\approx 1$ 1 nm and the migration processes of only five boundary hydrogen atoms
contribute to this factor. An increase in $L$ leads to the corresponding (proportional
to the number of boundary atoms, i.e., to $L$) increase in $A_{\textrm{AB}}$ and, hence,
to the increase in the probability of the formation of disordered configurations at a
given temperature and to the decrease in the onset time of their formation. However, the
statistical weight of these configurations remains small, about 10$^{-7}$ at room
temperature even at $L\sim 1 \mu$m. Thus, the interface (if it is a straight line) is
very stable against thermal disordering; for this reason, hybrid
graphene/graphane systems are promising for nanoelectronics.

Finally, note that only a zigzag graphene/graphane interface has been considered in this
work. It is also of interest to examine the thermal stability of an armchair interface.

This work was supported by the Russian Foundation for Basic Research (project no.
09-02-00701-a) and by the Ministry of Education and Science of the Russian Federation
(project no. 2.1.1/468, Federal Program {}``Development of the Scientific Potential of
Higher Education'').

\newpage

\centerline{Figure captions}

Fig. 1. C$_{88}$H$_{70}$ cluster as a model of the graphene/graphane system. The large and small
balls are carbon and hydrogen atoms, respectively.

Fig. 2. Typical configurations of the C$_{88}$H$_{70}$ cluster most often observed in the simulation
of the thermal stability of the graphene/graphane interface. The closed circles are carbon atoms.
The small open circles are passivating hydrogen atoms. The large open circles and squares are
hydrogen atoms located in different sublattices (above and below the cluster plane, respectively):
(A) the initial state, (B) the configuration formed after the migration of one hydrogen atom from
graphane to graphene by the distance of a C-C bond, (C) the configuration formed after the migration
of one hydrogen atom from graphane to graphene by the distance of two C-C bonds, and (D) the
configuration formed after the migration of each of the two hydrogen atoms belonging to different
sublattices from graphane to graphene by the distance of a C-C bond.

Fig. 3. Schematic profile of the potential energy $E_{\textrm{pot}}$ of the C$_{88}$H$_{70}$ cluster
near configurations {}``A'', {}``B'', and {}``D'' (see Fig. 2); S1 and S2 are the saddle points.
The energies are measured from the energy of configuration {}``A''. The heights of
the energy barriers are 
$U_{\textrm{AB}}=E_{\textrm{S1}}-E_{\textrm{A}}$,
$U_{\textrm{BA}}=E_{\textrm{S1}}-E_{\textrm{B}}$,
$U_{\textrm{BD}}=E_{\textrm{S2}}-E_{\textrm{B}}$, and
$U_{\textrm{DB}}=E_{\textrm{S2}}-E_{\textrm{D}}$.

\newpage

\vskip 20mm
\includegraphics[width=\hsize,height=14cm]{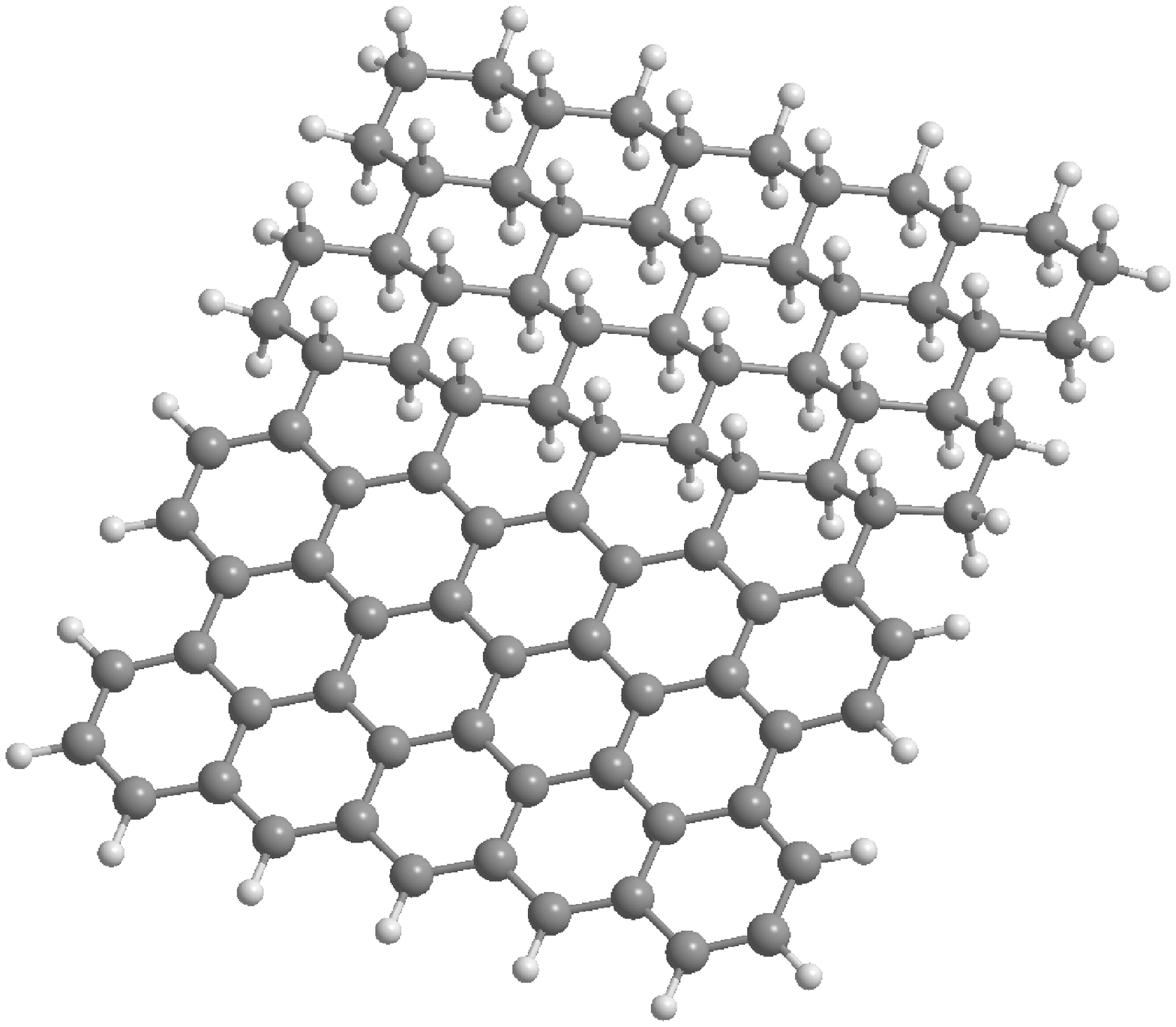}
\vskip 5mm
\centerline{Fig. 1}.

\newpage

\vskip 2mm
\includegraphics[width=\hsize,height=21.5cm]{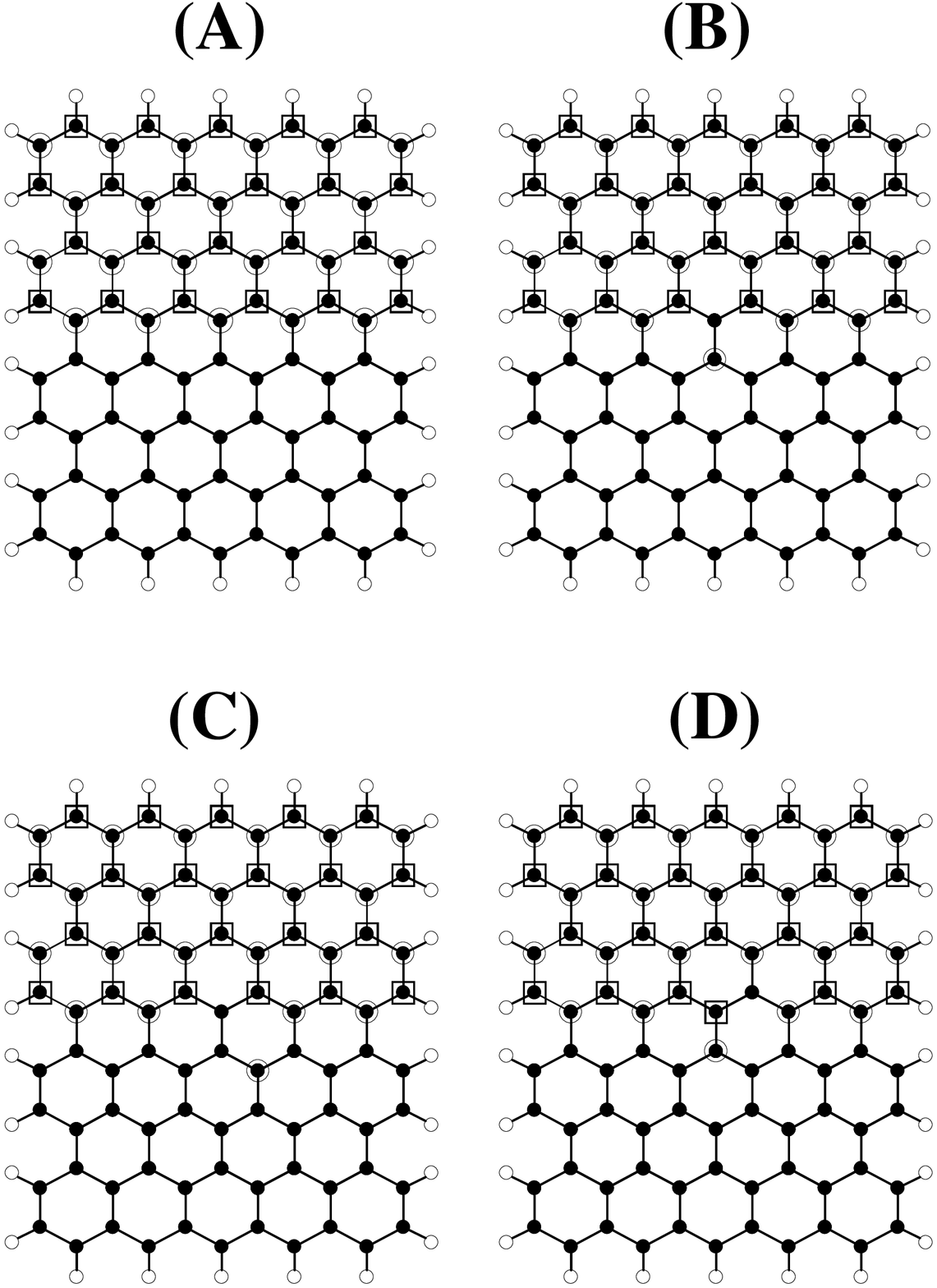}
\vskip 1mm
\centerline{Fig. 2}.

\newpage

\vskip 20mm
\includegraphics[width=\hsize,height=13cm]{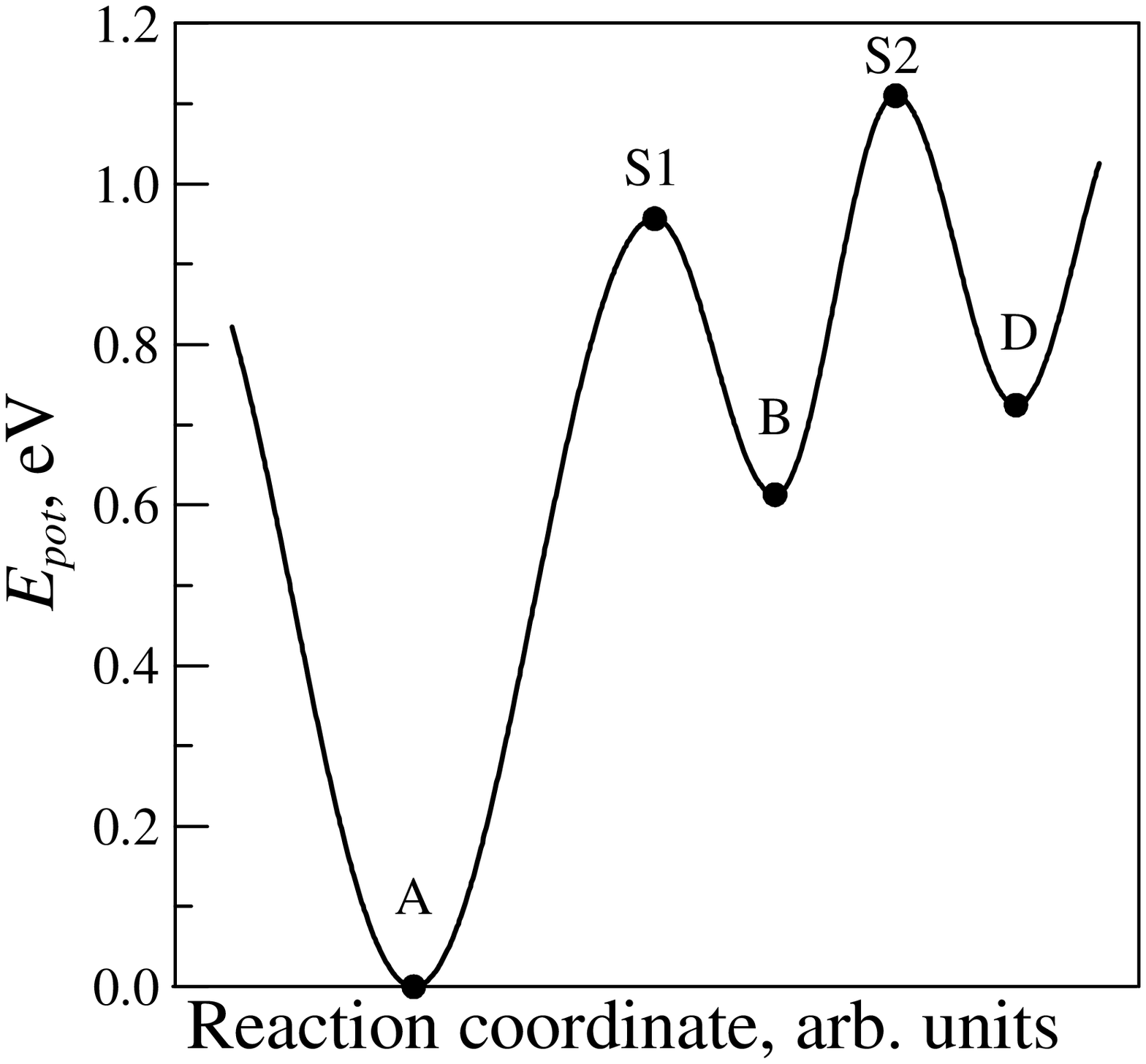}
\vskip 5mm
\centerline{Fig. 3}.

\end{document}